\documentclass[preprint,prc,aps,epsfig]{revtex4}
\usepackage{psfig}

\newcommand{\AmS}{{\protect\the\textfont2
  A\kern-.1667em\lower.5ex\hbox{M}\kern-.125emS}}
\def\beq{\begin{equation}}
\def\eeq{\end{equation}}
\def\beqa{\begin{eqnarray}}
\def\eeqa{\end{eqnarray}}
\def\MeV{\nobreak\,\mbox{MeV}}
\def\GeV{\nobreak\,\mbox{GeV}}

\begin{document}

\title{$J/\psi$-kaon cross section in meson exchange model}
\author{R.S. Azevedo and M. Nielsen}
\affiliation{Instituto de F\'{\i}sica, 
        Universidade de S\~{a}o Paulo, \\
        C.P. 66318,  05389-970 S\~{a}o Paulo, SP, Brazil}

\begin{abstract}
We calculate the cross section for the dissociation of $J/\psi$ by kaons
within the framework of a meson exchange model including anomalous
parity interactions. Off-shell effects at the vertices were handled 
with QCD sum rule estimates for the running coupling constants.
The total $J/\psi$-kaon cross section was found to be $1.0 \sim1.6$ mb for
$4.1\leq\sqrt{s}\leq5~\GeV$.
\end{abstract}

\maketitle

\section{Introduction}

Suppression of  $J/\psi$
in  relativistic  heavy  ion  collisions is still considered
as one of the important signatures  to  identify  the  possible
phase  transition  to  quark-gluon  plasma (QGP) \cite{ma86} (for a 
review of data  and interpretations see refs.~\cite{vo99,ge99}). 
Since there is no direct experimental information on $J/\psi$ absorption 
cross
sections by hadrons, several theoretical approaches have been proposed to
estimate their values.
In order to elaborate a theoretical description of the phenomenon, we have
first to choose the relevant degrees of freedom. Some
approaches were based on charm quark-antiquark dipoles interacting with the
gluons of a larger
(hadron target) dipole \cite{bhp,kha2,lo} or quark exchange between two
(hadronic) bags \cite{wongs,mbq}, or QCD sum rules \cite{nnmk,dlnn,dklnn}, 
whereas other works used the meson exchange mechanism
\cite{mamu98,haglin,linko,tho,haga,osl,nnr}.
In this case it is not easy to decide
in favor of quarks or hadrons because we are dealing with charm quark bound
states, which are small and massive enough to make perturbation theory
meaningful, but not small enough to make
non-perturbative effects negligible \cite{nnmk,dlnn,dklnn,dnn}.

The meson exchange approach was applied basically 
to $J/\psi-N$, $J/\psi-\pi$ and $J/\psi-\rho$ cross sections, with the 
only exception
of ref.~\cite{haglin} where $J/\psi-K$ cross section was also estimated.
However, as pointed out in ref.~\cite{linko}, there are some inconsistencies
in the Lagrangians defined in ref.~\cite{haglin}. In this work we 
extend the analysis done in ref.~\cite{aze} and
evaluate the $J/\psi-K$ cross section using a meson-exchange model,
considering anomalous parity terms as in ref.~\cite{osl}.

\section{Effective Lagrangians}

We follow refs.~\cite{haglin,linko,haga,osl} and start with the SU(4) 
Lagrangian
for the pseudo-scalar and vector mesons. The effective Lagrangians 
relevant for the study of the $J/\psi$ absorption by kaons are:
\beqa
{\cal L}_{K DD_s^*}&=&ig_{K DD_s^*}~ 
D_s^{* \mu}\left ( \bar D \partial_\mu \bar K - 
(\partial_\mu \bar D) \bar K \right ) + {\rm H.c.}~ , \label{kdds} 
\\
{\cal L}_{K D_sD^*}&=&ig_{K D_sD^*}~ 
D^{* \mu} \left ( \bar D_s \partial_\mu K - 
(\partial_\mu \bar D_s) K \right ) + {\rm H.c.}~ , \label{kdsd}
\\
{\cal L}_{\psi DD}&=&ig_{\psi DD}~ \psi^\mu 
\left ( D \partial_\mu \bar D -(\partial_\mu D) \bar D \right ) ~ ,
\label{jdd}\\ 
{\cal L}_{\psi D_sD_s}&=&ig_{\psi D_sD_s}~ \psi^\mu 
\left ( D_s \partial_\mu \bar D_s -(\partial_\mu D_s) \bar D_s \right ) ~ ,
\label{jdsds} \\
{\cal L}_{\psi D^*D^*}&=& ig_{\psi D^*D^*}~
\left [ \psi^\mu \left ( (\partial_\mu D^{* \nu}) \bar {D^*_\nu} 
- D^{* \nu} \partial_\mu \bar {D^*_\nu} \right )\right.\nonumber \\
&+&\left ( (\partial_\mu \psi^\nu) D^*_\nu -\psi^\nu 
\partial_\mu D^*_\nu 
\right )\bar {D^{* \mu}}
+\left. D^{* \mu} \left ( \psi^\nu \partial_\mu \bar {D^*_\nu} -
(\partial_\mu \psi^\nu) \bar {D^*_\nu} \right ) \right ] ~ , \label{jdede}\\ 
{\cal L}_{\psi D_s^*D_s^*}&=& ig_{\psi D_s^*D_s^*}~
\left [ \psi^\mu \left ( (\partial_\mu D_s^{* \nu}) {\bar D}_{s\nu}^* 
- D_s^{* \nu} \partial_\mu {\bar D}_{s\nu}^* \right )\right.\nonumber \\
&+&\left ( (\partial_\mu \psi^\nu) D_{s\nu}^* -\psi^\nu 
\partial_\mu D_{s\nu}^* 
\right ){\bar D}_s^{* \mu} 
+\left . D_s^{* \mu} \left ( \psi^\nu \partial_\mu {\bar D}_{s\nu}^* -
(\partial_\mu \psi^\nu) {\bar D}_{s\nu}^* \right ) \right ] ~ , 
\label{jdesdes}\\ 
{\cal L}_{\psi K D_sD^*}&=&-g_{\psi K  D_sD^*}~
\psi^\mu \left ( D^*_\mu K \bar D_s
+ D_s \bar K \bar {D^*_\mu} \right )  ~ , \label{kjdsd} \\ 
{\cal L}_{\psi K  DD_s^*}&=&-g_{\psi K  DD_s^*}~
\psi^\mu \left ( {\bar D}^*_{s\mu} K  D
+  \bar D \bar K  D^*_{s\mu} \right )  ~ , \label{kjdds}
\end{eqnarray}
where we have defined the charm meson and kaon iso-doublets $D\equiv(D^0,D^+)$,
$D^*\equiv(D^{*0},D^{*+})$ and $K\equiv(K^0,K^+)$. 

In addition to the normal terms given above, which were considered in 
\cite{aze}, there are also anomalous
parity terms introduced in ref.~\cite{osl} for the $J/\psi-\pi$ case.
In the $J/\psi-K$ case they are:
\begin{eqnarray}
&& {\cal L}_{\psi D^{*}D}=g_{\psi D^{*}D}\epsilon^{\mu \nu \alpha \beta}
\partial_{\mu}\psi_{\nu}\left(\partial_{\alpha}\bar{D}^{*}_{\beta}D+\partial_{
\alpha}D^{*}_{\beta}\bar{D}\right), \\
&& \mathcal{L}_{\psi D_{s}^{*}D_{s}}=g_{\psi D_{s}^{*}D_{s}}
\epsilon^{\mu \nu \alpha \beta}\partial_{\mu}\psi_{\nu}\left(\partial_{\alpha}
\bar{D}^{*}_{s\beta}D_{s}+\partial_{\alpha}D^{*}_{s\beta}\bar{D}_{s}\right), \\
&& \mathcal{L}_{K D_{s}^{*}D^{*}}=-g_{K D_{s}^{*}D^{*}}
\epsilon^{\mu \nu \alpha \beta}\left(\partial_{\mu}\bar{D}^{*}_{\nu}
\partial_{\alpha}D^{*}_{s\beta}\bar{K}+\partial_{\mu}D^{*}_{\nu}
\partial_{\alpha}
\bar{D}^{*}_{s\beta}K\right),\\
&& \mathcal{L}_{\psi K D_{s}D}=-ig_{K D_{s}D\psi}\epsilon^{\mu \nu \alpha 
\beta}\psi_{\mu}\left(\partial_{\nu}\bar{D}\partial_{\alpha}\bar{K}\partial_{
\beta}D_{s}-\partial_{\nu}D\partial_{\alpha}K\partial_{\beta}\bar{D}_{s}\right)
,\\
&& \mathcal{L}_{\psi K D_{s}^{*}D^{*}}=-ig_{K D^{*}_{s}D^{*}\psi}\epsilon^{
\mu \nu \alpha \beta}\psi_{\mu}\left(\bar{D}^{*}_{\nu}\partial_{\alpha}\bar{K}
D^{*}_{s\beta}-D^{*}_{\nu}\partial_{\alpha}K\bar{D}^{*}_{s\beta}\right) 
\nonumber \\
&&-ih_{\psi K D^{*}_{s}D^{*}}\epsilon^{\mu \nu \alpha \beta}\left(
\partial_{\mu}
D^{*}_{\beta}\psi_{\nu}\bar{D}^{*}_{s\alpha}K+\partial_{\mu}\bar{D}^{*}_{s
\alpha}\psi_{\nu}D^{*}_{\beta}K+3\partial_{\mu}\psi_{\nu}\bar{D}^{*}_{s
\alpha}D^{*}_{\beta}K\right. \nonumber \\
&&\left.-\partial_{\mu}\bar{D}^{*}_{\beta}\psi_{\nu}D^{*}_{s\alpha}\bar{K}-
\partial_{\mu}D^{*}_{s\alpha}\psi_{\nu}\bar{D}^{*}_{\beta}\bar{K}-3\partial_{
\mu}\psi_{\nu}D^{*}_{s\alpha}\bar{D}^{*}_{\beta}\bar{K}\right),
\end{eqnarray}

\begin{figure}[htb] \label{fig1}
\centerline{\psfig{figure=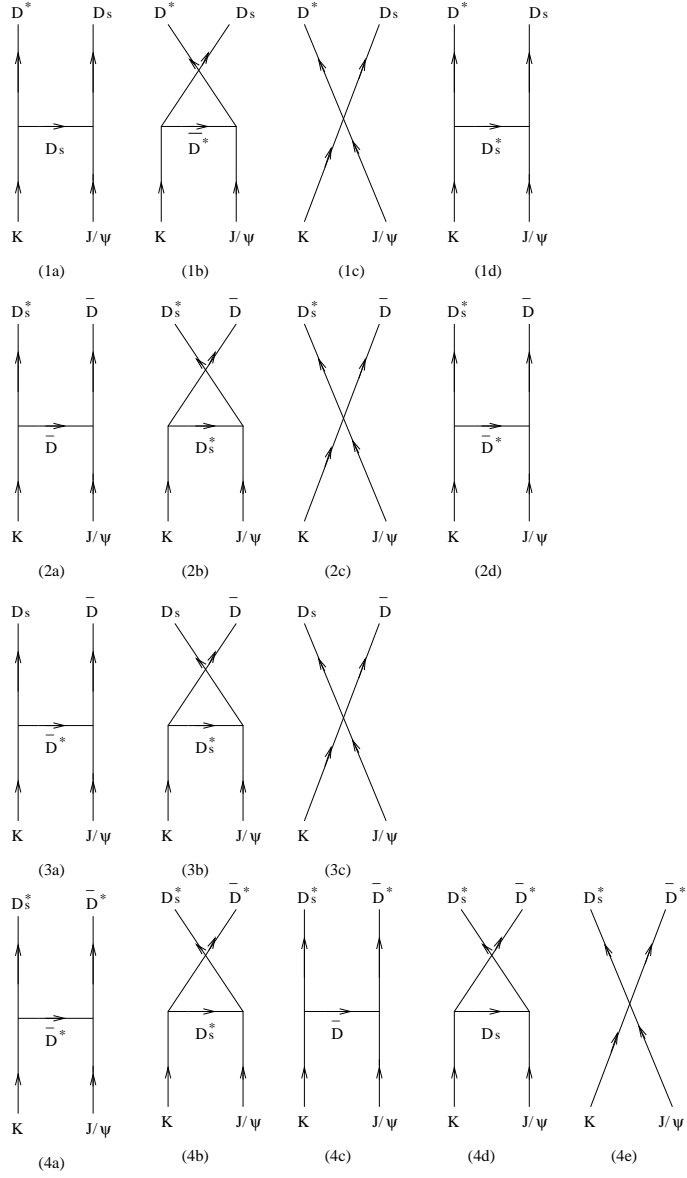,width=9cm,angle=0}}
\vspace{-.5cm}
\caption{\small{Diagrams for $J/\psi$ absorption processes:   
1) $K \psi \rightarrow D_s {\bar D}^*$;
2) $K \psi \rightarrow D_s^* \bar D$;
3) $K \psi \rightarrow D_s \bar D$;
4) $K \psi \rightarrow D_s^* {\bar D}^*$. Diagrams for
the processes $\bar K \psi \rightarrow {\bar D}_s D^*$,
$\bar K \psi \rightarrow {\bar D}_s D^*$, $\bar K \psi \rightarrow {\bar D}_s 
D$, and
$\bar K \psi \rightarrow {\bar D}_s^* D^*$
are similar to (1a)-(1d) through (4a)-(4d) respectively, but with each 
particle replaced by its anti-particle.}}  
\end{figure}

In Fig.~1 we show
the processes we want to study for the absorption of $J/\psi$ by kaons.
They  are:
\beqa
K J/\psi \rightarrow D_s {\bar D}^*, && \bar K J/\psi \rightarrow D^* 
{\bar D}_s,
\label{pro1}\\ 
K J/\psi \rightarrow D_s^* \bar D, && \bar K J/\psi \rightarrow D {\bar D}_s^*,
\label{pro2}\\
K J/\psi \rightarrow D_s \bar D, && \bar K J/\psi \rightarrow D {\bar D}_s.
\label{pro3}\\
K J/\psi \rightarrow D_s^* {\bar D}^*, && \bar K J/\psi \rightarrow D^* 
{\bar D}_s^*,
\label{pro4}
\eeqa
Since the two processes in eqs.~(\ref{pro1}), (\ref{pro2}), (\ref{pro3}) and 
(\ref{pro4}) have the same 
cross section, in Fig.~1 we only show the diagrams for the first process
in eqs.~(\ref{pro1}) through (\ref{pro4}).

Defining  the four-momentum of the kaon and 
the $J/\psi$  by $p_1$, $p_2$ respectively, and the four-momentum of the
vector and pseudo-scalar final-state mesons
respectively by $p_3$ and $p_4$, the full amplitude for the 
processes $K \psi \rightarrow D_s {\bar D}^*$ and $K \psi \rightarrow D_s^* 
\bar D$, shown in diagrams (1) and (2) of Fig.~1, 
without isospin factors and before summing and averaging over external 
spins, is given by

\begin{eqnarray}
{\cal M}_i \equiv {\cal M}_i^{\nu \lambda} 
~\epsilon_{2 \nu} \epsilon_{3 \lambda}^* 
=\left ( \sum_{j=a,b,c,d} {\cal M}_{ij}^{\nu \lambda} \right )
\epsilon_{2 \nu} \epsilon_{3 \lambda}^*, \mbox{ for }i=1,2 
\label{m1}
\end{eqnarray}
with
\begin{eqnarray}
{\cal M}_{1a}^{\nu \lambda}&=& -g_{K D_s D^*} g_{\psi D_s D_s}~
(-2p_1+p_3)^\lambda \left (\frac{1}{t-m_{D_s}^2} \right )
(p_1-p_3+p_4)^\nu, \nonumber \\
{\cal M}_{1b}^{\nu \lambda}&=& g_{K D_s D^*} g_{\psi D^* D^*}~
(-p_1-p_4)^\alpha \left ( \frac{1}{u-m_{D^*}^2} \right )
\left [ g_{\alpha \beta}-\frac{(p_1-p_4)_\alpha (p_1-p_4)_\beta}{m_{D^*}^2}
\right ]\nonumber \\
&\times&
 \left [ (-p_2-p_3)^\beta g^{\nu \lambda}
+(-p_1+p_2+p_4)^\lambda g^{\beta \nu}
+(p_1+p_3-p_4)^\nu g^{\beta \lambda} \right ] , \nonumber \\
{\cal M}_{1c}^{\nu \lambda}&=& -g_{\psi K D_s D^*}~ g^{\nu \lambda},
\nonumber \\
{\cal M}_{1d}^{\nu \lambda} &=& -\frac{g_{KD_{s}^{*}D^{*}}g_{\psi 
D_{s}^{*}D_{s}}}
{t-m_{\bar{D}^{*}_{s}}^{2}}\epsilon^{\lambda \rho \sigma \alpha}\epsilon^
{\nu \gamma \delta \beta}\left [g_{\alpha \beta} - \frac{(p_{1} - 
p_{3})_{\alpha}
(p_{1} - p_{3})_{\beta}}{m_{\bar{D}_{s}^{*}}^{2}}\right ]p_{1\sigma}p_{2\gamma}
p_{3\rho}p_{4\delta}.
\label{pij1}
\end{eqnarray}
\begin{eqnarray}
{\cal M}_{2a}^{\nu \lambda}&=& -g_{K D D_s^*} g_{\psi D D}~
(-2p_1+p_3)^\lambda \left (\frac{1}{t-m_{D}^2} \right )
(p_1-p_3+p_4)^\nu, \nonumber \\
{\cal M}_{2b}^{\nu \lambda}&=& g_{K D D_s^*} g_{\psi D_s^* D_s^*}~
(-p_1-p_4)^\alpha \left ( \frac{1}{u-m_{D_s^*}^2} \right )
\left [ g_{\alpha \beta}-\frac{(p_1-p_4)_\alpha (p_1-p_4)_\beta}{m_{D_s^*}^2}
\right ] 
\nonumber \\
&\times&
\left [ (-p_2-p_3)^\beta g^{\nu \lambda}+
(-p_1+p_2+p_4)^\lambda g^{\beta \nu}
+(p_1+p_3-p_4)^\nu g^{\beta \lambda} \right ] , \nonumber \\
{\cal M}_{2c}^{\nu \lambda}&=& -g_{\psi K D D_s^*}~ g^{\nu \lambda} , 
\nonumber \\
{\cal M}_{2d}^{\nu \lambda} &=& -\frac{g_{KD_{s}^{*}D^{*}}g_{\psi D^{*}D}}{t-m_
{\bar{D}^{*}}^{2}}\epsilon^{\lambda \rho \sigma \alpha}\epsilon^{\nu \gamma 
\delta \beta}\left [ g_{\alpha \beta} - \frac{(p_{1} - p_{3})_{\alpha}(p_{1} 
- p_{3})_{\beta}}{m_{\bar{D}^{*}}^{2}}\right ]p_{1\sigma}p_{2\gamma}p_{3\rho}
p_{4\delta},
\label{pij2}
\end{eqnarray}
where $t=(p_1-p_3)^2$ and $u=(p_1-p_4)^2$.

We can see that the differences between these two processes are basically due
to the meson exchanged.
It can be shown \cite{linko,osl} that the full amplitudes 
${\cal M}_{i}^{\nu \lambda}$ (for $i=1,2$) given above satisfy current 
conservation: ${\cal M}_{i}^{\nu \lambda}p_{2\nu}={\cal M}_{i}^{\nu \lambda}
p_{3\lambda}=0$.

Calling the four-momentum of the two pseudo-scalar final-state mesons
respectively by $p_3$ and $p_4$, the full amplitude for the 
processes $K \psi \rightarrow D_s {\bar D}$ shown in diagram (3) of Fig.~1 is 
\begin{equation}
{\cal M}_{3} \equiv {\cal M}_{3}^{\nu}\epsilon_{2\nu}=
\left(\sum_{i=a,b,c}{\cal M}_{3i}^{\nu}\right)
\epsilon_{2\nu},
\end{equation}
with
\begin{eqnarray}
&& {\cal M}^{\nu }_{3a} = \frac{g_{KD_{s}D^{*}}g_{\psi D^{*}D}}
{t-m_{\bar{D}^{*}}
^{2}}\epsilon^{\nu \beta \gamma \delta}\left(g_{\alpha \beta} - 
\frac{(p_{1} - 
p_{3})_{\alpha}(p_{1} - p_{3})_{\beta}}{m_{\bar{D}^{*}}^{2}}\right)
(p_{1}+p_{3})^{
\alpha}p_{2\gamma}p_{4\delta}, \nonumber \\
&& {\cal M}^{\nu}_{3b} = -\frac{g_{KDD_{s}^{*}}g_{\psi D_{s}^{*}D_{s}}}
{u-m_{D_{s}
^{*}}^{2}}\epsilon^{\nu \beta \gamma \delta}\left(g_{\alpha \beta} - 
\frac{(p_{3} 
- p_{2})_{\alpha}(p_{3} - p_{2})_{\beta}}{m_{D_{s}^{*}}^{2}}\right)
(p_{1}+p_{4})^
{\alpha}p_{2\gamma}p_{3\delta},  \nonumber \\ 
&& {\cal M}^{\nu}_{3c} = -g_{\psi KDD_{s}}\epsilon^{\nu \delta \lambda \gamma}
p_{1\delta}p_{3\lambda}p_{4\gamma}.
\label{pij3}
\end{eqnarray}

For the diagram (4) in Fig.~1, representing the processes 
$K \psi \rightarrow D_s^* {\bar D}^*$, calling the four-momentum of the 
two vector  final-state mesons
respectively by $p_3$ and $p_4$, the full amplitude is given by
\begin{equation}
{\cal M}_{4} \equiv {\cal M}_{4}^{\nu \lambda \mu}\epsilon_{2\nu} 
\epsilon_{3\lambda}^{*}\epsilon_{4\mu}^{*}=\left(\sum_{i=a,b,c,d,e}
{\cal M}_{4i}^{\nu \lambda \mu}\right)
\epsilon_{2\nu} \epsilon_{3\lambda}^{*}\epsilon_{4\mu}^{*}, 
\end{equation}
with
\begin{eqnarray}
&& {\cal M}_{4a}^{\nu \lambda \mu} =-g_{KD^{*}D_{s}^{*}}g_{\psi
D^{*}D^{*}}\frac{1}{t - m_{\bar{D}^{*}}^{2}}\left(g_{\alpha \beta} - 
\frac{(p_{2} - p_{4})_{\alpha}(p_{2} - p_{4})_{\beta}}{m_{\bar{D}^{*}}^{2}}
\right)
\epsilon^{\lambda \sigma \rho \alpha}p_{3\sigma}p_{1\rho}[(2p_{2} 
- p_{4})^{\mu}g^{\beta \nu}  \nonumber \\
&& + (-p_{2} - p_{4})^{\beta}g^{\mu \nu} + (- p_{2} +2p_{4})^{\nu}g^{\beta 
\mu}], \nonumber \\
&& {\cal M}_{4b}^{\nu \lambda \mu}=-g_{KD^{*}D_{s}^{*}}g_{\psi
D_{s}^{*}D_{s}^{*}}\frac{1}{u - m_{D_{s}^{*}}^{2}}\left(g_{\alpha \beta} - 
\frac{(p_{1} - p_{4})_{\alpha}(p_{1} - p_{4})_{\beta}}{m_{D_{s}^{*}}^{2}}
\right)
\epsilon^{\mu \sigma \rho \alpha}p_{1\sigma}p_{4\rho}[(2p_{3} - p_{2})^{
\nu}g^{\beta \lambda}  \nonumber   \\
&& + (-p_{2} - p_{3})^{\beta}g^{\nu \lambda} + (-p_{3} + 2p_{2})^{\lambda}
g^{\beta \nu}], \nonumber \\
&&  {\cal M}_{4c}^{\nu \lambda \mu}=g_{KDD_{s}^{*}}g_{\psi
D{*}D}\frac{1}{t - m_{\bar{D}}^{2}}\epsilon^{\nu \mu \gamma \delta}(2p_{1}
-p_{3})^{\lambda}p_{2\gamma}p_{4\delta}, \nonumber \\
&&{\cal M}_{4d}^{\nu \lambda \mu}=-g_{KD_{s}D^{*}}g_{\psi
D_{s}{*}D_{s}}\frac{1}{u - m_{D_{s}}^{2}}\epsilon^{\nu \lambda \gamma 
\sigma}(2p_{1}-p_{4})^{\mu}p_{2\gamma}p_{3\sigma}, \nonumber \\
&&{\cal M}_{4e}^{\nu \lambda \mu}=-g_{\psi KD^{*}D_{s}^{*}}
\epsilon^{\nu \lambda \mu \sigma}p_{1\sigma} + h_{\psi KD^{*}D_{s}^{*}}
\epsilon^{\nu \lambda \mu \sigma}(p_{4}+p_{3}-3p_{2})_{\sigma}.
\label{pij4}
\end{eqnarray}

After averaging (summing) over initial (final) spins 
and including isospin factors, the cross sections 
for these four processes are given by 
\beqa\label{jkaon12}
\frac {d\sigma_i}{dt}&=& \frac {1}{192 \pi s p_{0,\rm cm}^2} 
{\cal M}_i^{\nu \lambda} {\cal M}_i^{*\nu^\prime \lambda^\prime}
\left ( g_{\nu \nu^\prime}-\frac{p_{2 \nu} p_{2 \nu^\prime}} {m_2^2} \right )
\left ( g_{\lambda \lambda^\prime}
-\frac{p_{3 \lambda} p_{3 \lambda^\prime}} {m_3^2} \right ), \mbox{ for }i=1,2
\end{eqnarray}
\beq\label{jkaon3}
\frac {d\sigma_3}{dt}= \frac {1}{192 \pi s p_{0,\rm cm}^2} 
{\cal M}_3^{\nu} {\cal M}_3^{*\nu^\prime}
\left ( g_{\nu \nu^\prime}-\frac{p_{2 \nu} p_{2 \nu^\prime}} {m_2^2} \right )
\eeq
and
\beq\label{jkaon14}
\frac {d\sigma_4}{dt}= \frac {1}{192 \pi s p_{0,\rm cm}^2} 
{\cal M}_4^{\nu \lambda\mu} {\cal M}_4^{*\nu^\prime \lambda^\prime\mu^\prime }
\left ( g_{\nu \nu^\prime}-\frac{p_{2 \nu} p_{2 \nu^\prime}} {m_2^2} \right )
\left ( g_{\lambda \lambda^\prime}
-\frac{p_{3 \lambda} p_{3 \lambda^\prime}} {m_3^2} \right )
\left ( g_{\mu \mu^\prime}
-\frac{p_{4 \mu} p_{4 \mu^\prime}} {m_4^2} \right ), 
\eeq
with $s=(p_1+p_2)^2$, and  
\begin{eqnarray}
p_{0,\rm cm}^2=\frac {\left [ s-(m_1+m_2)^2 \right ]
\left [ s-(m_1-m_2)^2 \right ]}{4s}
\end{eqnarray}
being the squared three-momentum of initial-state mesons in the 
center-of-momentum (c.m.) frame. 

\section{Numerical results}
\subsection{Coupling constants and form factors}

To estimate the cross sections we have first to determine the coupling 
constants of the effective Lagrangians.
Exact SU(4) symmetry 
would give the following relations among the coupling constants 
\cite{linko,osl}:
\begin{eqnarray}
&&g_{K D_sD^*}=g_{K DD_s^*}=\frac{g}{2\sqrt{2}}~, \nonumber \\
&&g_{\psi DD}=g_{\psi D_sD_s}=g_{\psi D^* D^*}=g_{\psi D_s^* D_s^*}=
\frac{g}{\sqrt 6}~, \nonumber \\
&&g_{\psi K D_sD^*}=g_{\psi K DD_s^*}=\frac{g^2}{4 \sqrt 3}~,\nonumber \\
&& g_{\psi D^{*}D}=g_{\psi D_{s}^{*}D_{s}}={g_{KD_{s}^{*}D^{*}}\over\sqrt{3}}
={\sqrt{2}g_a^2N_c\over 64\sqrt{3}\pi^2F_\pi}
, \nonumber \\
&&g_{\psi KDD_{s}}=\frac{g_aN_c}{96\sqrt{6}\pi^2F_\pi^3}, \nonumber \\
&&g_{\psi KD^{*}D_{s}^{*}}=5h_{\psi K D^{*}D_{s}^{*}}=
\frac{5g_a^3N_c}{2^8\sqrt{3}\pi^2F_\pi}, 
\label{su4}
\end{eqnarray}
where $N_c=3$ and $F_\pi=132\MeV$.

Since none of the above couplings are known experimentally, one has to use
models to estimate them. In ref.~\cite{aze} we have calculated the cross 
section for the processes
$J/\psi~\mbox{kaons}\rightarrow$ $\bar{D}^*~D_s + D^*~\bar{D}_s +\bar{D}~
D_s^* +\bar{D}_s^*~D$  considering only the normal terms given by
the Lagrangians in Eqs.~(\ref{kdds}) through (\ref{kjdds}), and using
two different ways to estimate the coupling constants: vector meson dominance
model
estimate of $g_{\psi DD}$ plus SU(4) relations, and the experimental 
result for $g_{\rho\pi\pi}$ plus SU(4) relations. We showed that the results
for the cross section can vary by almost one order of magnitude,  
depending on the values of the coupling constants used,
even without considering form factors in the
hadronic vertices \cite{linko,osl}. This gives an idea of how important it is
to have a good estimate of the value of the coupling constants. In a recent
work \cite{haga2}, the $J/\psi-\pi$ and $J/\psi-\rho$ cross sections
were evaluated by using form factors and coupling constants estimated 
using QCD sum rules \cite{nos1,nos2,nos3,nos4}. The results in 
ref.~\cite{haga2} show that,
with  appropriate form factors, even the behavior of the cross section
as a function of $\sqrt{s}$ can change. In this work
we use the form factors in the vertices ${J/\psi DD}$, ${J/\psi D^{*}D}$
and $\pi D^*D$, determined from QCD sum rules \cite{nos2,nos5}, and the 
above SU(4) relations to 
estimate the form factors and coupling constants in all vertices.

From ref.~\cite{nos5} we get $g_{\psi DD^{*}}=4.0~\GeV^{-1}$ and
$g_{\psi DD}=5.8$. Using these numbers in the SU(4) relations given in
Eq.~(\ref{su4}) we obtain
\begin{eqnarray}
&&g_{\psi DD}=g_{\psi D_sD_s}=g_{\psi D^* D^*}=g_{\psi D_s^* D_s^*}=
5.8,\;\;\;
g_{K D_sD^*}=g_{K DD_s^*}=5.0,\nonumber \\
&&g_{\psi K D_sD^*}=g_{\psi K DD_s^*}=
28.8,\;\;\;g_{\psi D^{*}D}=g_{\psi D_{s}^{*}D_{s}}=4.0\GeV^{-1},
\;\;\; g_{KD_{s}^{*}D^{*}}=7.0\GeV^{-1}, \nonumber \\
&&g_{\psi KDD_{s}}=6.6\GeV^{-3},\;\;\;g_{\psi KD^{*}D_{s}^{*}}=41.6\GeV^{-1}, 
\;\;\; h_{\psi K D^{*}D_{s}^{*}}=8.3\GeV^{-1}.
\label{sun}
\end{eqnarray}

The form factors given in ref.~\cite{nos5} are
\beq
g^{(D^{*})}_{\psi DD^{*}}(t)=g_{\psi DD^{*}}
\left(5~e^{-\left(\frac{27-t}{18.6}\right)^2}\right)=g_{\psi DD^{*}}~h_1(t),
\label{h1}
\eeq
\beq
g^{(D)}_{\psi DD^{*}}(t)=g_{\psi DD^{*}}\left(3.3
~e^{-\left(\frac{26-t}{21.2}\right)^2}\right)=g_{\psi DD^{*}}~h_2(t),
\label{h2}
\eeq
\beq
g^{(D)}_{\psi DD}(t)=g_{\psi DD}
\left(2.6~e^{-\left(\frac{20-t}{15.8}\right)^2}\right)=g_{\psi DD}~h_3(t),
\label{h3}
\eeq
where $g_{123}^{(1)}$ means the form factor at the vertex involving the
mesons $123$ with the meson $1$ off-shell. In the above equations the
numbers in the exponentials are in units of $\GeV^2$.
Since there is no QCD sum rule calculation for the form factors at the
vertices ${KD^*_s D}$ or $KD^*D_s$, we make the supposition that they are
similar to the form factor at the vertex $\pi D^*D$. From ref.~\cite{nos2}
we get
\beq
g^{(D)}_{\pi D^*D}(t)=g_{\pi D^*D}\left({(3.5\GeV)^2-m_D^2\over (3.5\GeV)^2-t}
\right)=g_{\pi D^*D}~h_4(t,m_D^2).
\label{h4}
\eeq

With these form factors the amplitudes will be modified in the following
way: 
\beq
{\cal M}_{ia}\rightarrow h_3(t)h_4(t,m_{ia}^2){\cal M}_{ia},\;\;\;
{\cal M}_{ib}\rightarrow h_3(u)h_4(u,m_{ib}^2){\cal M}_{ib},
\label{ff1}
\eeq
for $i=1,2\mbox{ and }4$, with $m_{1a}=m_{D_s},~m_{1b}=m_{4a}=m_{D^*},~
m_{2a}=m_D,$ and $m_{2b}=m_{4b}=m_{D_s^*}$.
\beq
{\cal M}_{ic}\rightarrow {1\over2}\left(h_3(t)h_4(t,m_{ia}^2)+h_3(u)
h_4(u,m_{ib}^2)\right){\cal M}_{ic},\;\;\;
{\cal M}_{id}\rightarrow h_1(t)h_4(t,m_{id}^2){\cal M}_{id},
\eeq
for $i=1,2$ with $m_{1d}=m_{D_s^*}$ and $m_{2d}=m_{D^*}$.
\beqa
{\cal M}_{3a}\rightarrow h_1(t)h_4(t,m_{D^*}^2){\cal M}_{3a},&&\;\;\;
{\cal M}_{3b}\rightarrow h_1(u)h_4(u,m_{D_s^*}^2){\cal M}_{3b}\nonumber\\
{\cal M}_{3c}\rightarrow {1\over2}\left(h_1(t)h_4(t,m_{D^*}^2)\right.&+&
\left.h_1(u)h_4(u,m_{D_s^*}^2)\right){\cal M}_{3c},
\eeqa
and
\beqa
{\cal M}_{4c}&\rightarrow& h_2(t)h_4(t,m_{D}^2){\cal M}_{4c},\;\;\;
{\cal M}_{4d}\rightarrow h_2(u)h_4(u,m_{D_s}^2){\cal M}_{4d},\;\;\;
{\cal M}_{4e}\rightarrow {1\over4}\left(h_3(t)h_4(t,m_{D^*}^2)\right.
\nonumber\\
&+&\left.h_3(u)
h_4(u,m_{D_s^*}^2)+h_2(t)h_4(t,m_{D}^2)+h_2(u)
h_4(u,m_{D_s}^2)\right){\cal M}_{4e}.
\label{ff2}
\eeqa

\subsection{Cross sections}

We first give the cross sections for the $J/\psi$ absorption by kaons without
considering the form factors, {\it i.e.}, we use the expressions for the 
amplitudes in Eqs.~(\ref{pij1}) through (\ref{pij4}). We will be always
including the contributions for both $J/\psi K$ and $J/\psi\bar{K}$.
\begin{figure}[htb]
\centerline{\psfig{figure=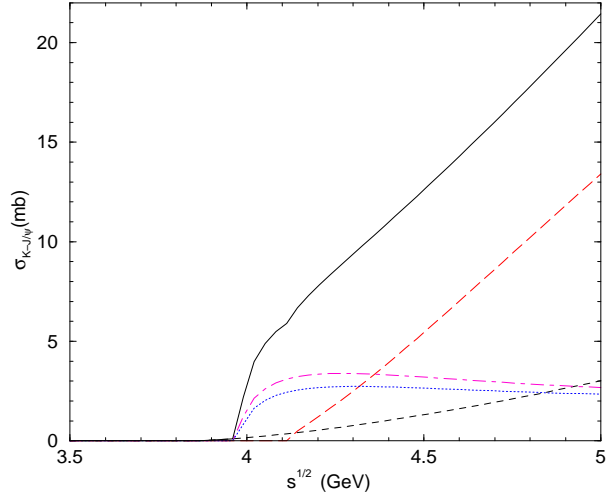,width=8cm}}
\vspace{-.5cm}
\caption{\small{Total cross sections, without form factors,
 for the processes $J/\psi~\mbox{kaons}
\rightarrow$ $\bar{D}^*~D_s + D^*~\bar{D}_s$ (dot-dashed line), $\bar{D}~
D_s^* +\bar{D}_s^*~D$ (dotted line), $\bar{D}~D_s + D~\bar{D}_s$ (dashed line)
and $\bar{D}^*~D_s^* + D^*~\bar{D}_s^*$ (long-dashed line).
The solid line gives the total $J/\psi$ dissociation by kaons cross section.}}
\label{fig2}
\end{figure}

In Fig. 2 we show the cross section of $J/\psi$ dissociation by kaons as a 
function of the initial  energy $\sqrt{s}$. 
The dot-dashed, dotted, dashed, long-dashed and solid lines give the
contributions for the processes $J/\psi~\mbox{kaons}
\rightarrow$ $\bar{D}^*~D_s + D^*~\bar{D}_s$, $\bar{D}~
D_s^* +\bar{D}_s^*~D$, $\bar{D}~D_s + D~\bar{D}_s$,
$\bar{D}^*~D_s^* + D^*~\bar{D}_s^*$ and total respectively. We see that the
for $\sqrt{s}>4.4~\GeV$ the process $J/\psi~\mbox{kaons}
\rightarrow$ $\bar{D}^*~D_s^* + D^*~\bar{D}_s^*$ dominates. However,
for smaller values of $\sqrt{s}$ the processes given by diagrams (1) and (2) 
in Fig.~1 are the most important ones. This is similar to what was
found in ref.~\cite{osl} for the $J/\psi$ dissociation by pions.

\begin{figure}[htb]
\centerline{\psfig{figure=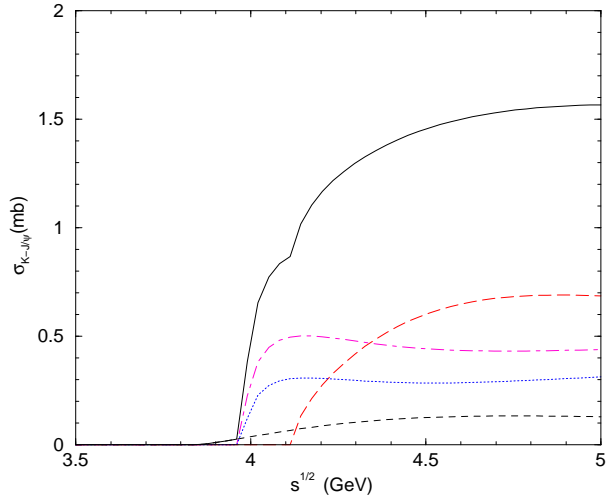,width=8cm}}
\vspace{-.5cm}
\protect\caption{\small{Same as in Fig.~2 but with form factors.}}
\label{fig3}
\end{figure}
In Fig.~3 we show the same processes considered in Fig.~2, but with
form factors. This means that we are now using the amplitudes given by 
Eqs.~(\ref{ff1}) through (\ref{ff2}). The first important conclusion
is that the use of appropriate form factors do change the behavior
of the cross section as a function of $\sqrt{s}$, as obtained in \cite{haga2}.
The processes more affected by this change are the ones
represented by diagrams (3) and (4) in Fig.~1.
While the total cross section obtained without form factors show a very strong
grown with $\sqrt{s}$. This is not more the case when the total
cross section is obtained with form factors, as can be seen in Fig.~4,
where we show  the total cross section evaluated with
and without form factors.
\begin{figure}[htb]
\centerline{\psfig{figure=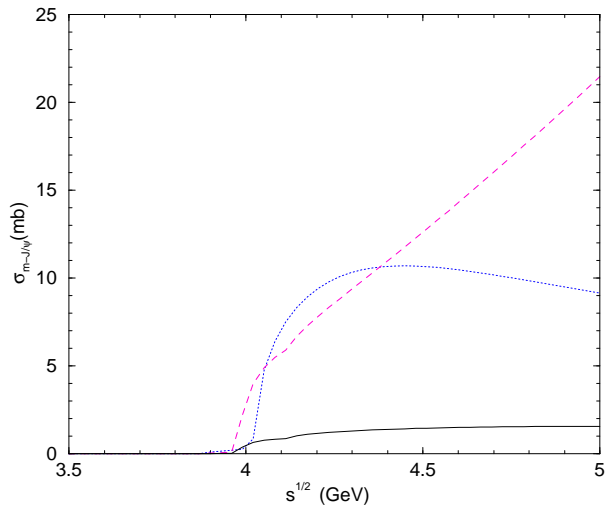,width=8cm}}
\vspace{-.5cm}
\protect\caption{\small{Total $J/\psi$  absorption cross section 
as a function of the initial energy. The solid and dashed lines give the
results for  $J/\psi$  absorption by kaons
with and without form factors respectively. The dotted line gives the results
for  $J/\psi$  absorption by pions with form factors.}}
\label{fig4}
\end{figure}

Other important result of our calculation is the fact that, using
appropriate form factors with cut-offs of order of $\sim4~\GeV$ 
(see Eqs.(\ref{h1})
through (\ref{h4})), the value of the cross section  can be reduced
by one order of magnitude. The same effect was obtained in 
refs.~\cite{linko,osl} using monopole form factors, but with cut-offs of 
order of $\sim1~\GeV$, which are considered very small for charmed mesons.

In Fig.~4 we also show, for comparison, 
the total cross section for $J/\psi$  absorption by 
pions (dotted line)
using the same form factors and coupling constants given here, and
the experimental value for the $D^*D\pi$ coupling constant:
$g_{\pi D^*D}= 12.6$ \cite{cleo}.
It is important to mention that the smallness of the value of the
total $J/\psi$-kaon absorption cross section, as compared with the
$J/\psi$-pion absorption cross section, is due to
the use of the experimental value for $g_{\pi D^*D}$, which is much bigger
than what one would get by using SU(4) relations: $g_{\pi D^*D}=g/4=3.6$. 
In ref.~\cite{aze} we have
showed that, using coupling constants related by SU(4) relations,
the $J/\psi$-kaon absorption 
cross section is even bigger than the $J/\psi$-pion absorption cross section.
Therefore, once more this shows how important it is to have good
estimates for the couplings.

\section{Conclusions}

We have studied the cross section of $J/\psi$ dissociation 
by kaons in a meson-exchange model that
includes pseudo-scalar-pseudo-scalar-vector-meson couplings,
three-vector-meson couplings, pseudo-scalar-vector-vector-meson couplings
and four-point couplings. Off-shell effects at the vertices were handled 
with QCD sum rule estimates for the form factors. The inclusion of anomalous 
parity interactions (pseudo-scalar-vector-vector-meson couplings) has opened
additional channels to the absorption mechanism. Their contribution
are very important especially for large values of the initial energy,
$\sqrt{s}>4.4~\GeV$.

As shown in Fig.~2 our results, without form factors, have the same energy 
dependence of $J/\psi$ absorption by pions from ref.~\cite{osl}. The
inclusion of the form factors changes the energy dependence of the
absorption cross section in a nontrivial way, as shown in Fig.~3. This 
modification in the energy dependence is
similar to what was found in ref.~\cite{haga2} for $J/\psi$ absorption by 
pions.

With QCD sum rules estimates for the coupling constants and form factors,
the total $J/\psi$-kaon cross section was found to be $1.0 \sim1.6$ mb for
$4.1\leq\sqrt{s}\leq5~\GeV$. Using the same form factors and the experimental
value for $g_{\pi D^*D}$ we get for the $J/\psi$-pion total dissociation
cross section $9.0 \sim10.0$ mb, in the same energy range.

\section*{Acknowledgments}

This work was supported by CNPq and FAPESP.

\end{document}